\pdfoutput=1
\documentclass[runningheads]{llncs}
\usepackage[T1]{fontenc}

\usepackage{graphicx}

\newcommand\socialbot{social bot}
\newcommand\Socialbot{Social bot}
\newcommand\socialbots{social bots}
\newcommand\Socialbots{Social bots}
\newcommand\Botometer{{\it Botometer}}
\newcommand\BotOrNot{{\it BotOrNot}}

\usepackage{moreverb,url}

\usepackage[colorlinks,bookmarksopen,bookmarksnumbered,citecolor=red,urlcolor=red]{hyperref}

\newcommand\BibTeX{{\rmfamily B\kern-.05em \textsc{i\kern-.025em b}\kern-.08em
T\kern-.1667em\lower.7ex\hbox{E}\kern-.125emX}}

\begin{document}


\title{Investigating the Validity of Botometer-based Social Bot Studies}

\author{F. Gallwitz\inst{1} and M. Kreil\inst{2}}

\institute{Nuremberg Institute of Technology, Kesslerplatz 10, 90486
Nuremberg, Germany \email{florian.gallwitz@th-nuernberg.de} \and SWRdata, Hans-Bredow-Strasse 9,
76530 Baden-Baden, Germany \email{contact1@michael-kreil.de}}


\maketitle

\begin{abstract} 
	The idea that social media platforms like Twitter are inhabited by vast
	numbers of \socialbots\ has become widely accepted in recent
	years. \Socialbots\ are
	assumed to be automated social media accounts
	operated by malicious actors with the goal of  manipulating public
	opinion. They are credited with the ability to produce content
	autonomously and to interact with human users. \Socialbot\ activity has
	been reported in many different political contexts, including
	the
	U.S. presidential elections, discussions about migration, climate
	change, and COVID-19.
	However, the relevant publications
	either use crude and questionable heuristics to discriminate between supposed 
	\socialbots\ and humans or---in the vast majority of the cases---fully 
	rely on the output of automatic bot detection tools, most
	commonly \Botometer. 
        In this paper, we point out a fundamental
	theoretical flaw in the widely-used study design for
	estimating the prevalence of \socialbots. 
	Furthermore, we empirically investigate
	the validity of peer-reviewed \Botometer-based studies by
	closely and systematically inspecting hundreds of accounts
	that had been counted as \socialbots. We
 	were unable to find a single \socialbot. Instead, we
	found mostly accounts undoubtedly operated by human users, the vast majority of them
	using Twitter in an inconspicuous and unremarkable fashion
	without the slightest traces of automation. We conclude that
	studies claiming to investigate the prevalence, properties, or
	influence of social bots based on \Botometer\ have, in reality,
	just investigated false positives and artifacts of this approach.
\end{abstract}


\section{Introduction}
\Socialbot\ or not? An extensive amount of research has been published in recent years suggesting 
that social media platform like Twitter are inhabited by vast numbers of \socialbots. These are supposed to be  accounts pretending to be human users but which are operated automatically by malicious actors with the goal of manipulating public opinion. The supposed influence of \socialbots\ in political discussions has raised significant concerns, particularly given their alleged potential to adversely impact democratic outcomes. 

The idea of \socialbot\ armies has been widely and frequently covered by media outlets across the world, with new reports of supposed \socialbot\ activity appearing almost on a weekly basis for the last couple of years, in the context of a wide variety of different topics. Discussions that have reportedly been attacked by \socialbot\ activity include the Brexit referendum, elections and political unrests in various countries, climate change, immigration, racial unrest, cannabis, vaping, COVID-19, vaccines, and even celebrity gossip.  As a consequence, political countermeasures against the supposed dangers of \socialbot\ activity have been discussed and legal regulations have been implemented, for example California's Bot Disclosure Law (2019) or Germany's \lq Medienstaatsvertrag\rq\ (2020).

Many of these news reports and also most scientific publications about \socialbots\ from research groups around the
world are based on \Botometer\ (originally called \BotOrNot), which has often been referred to
as the ``state-of-the-art bot detection method''.
A Google Scholar search in May 2022 using
the query  \texttt{"BotOrNot" OR "Botometer"} returns 1,720 results.

A typical definition of a \socialbot\ is given in~\cite{allem2018could}:
``{\it Social bots are automated accounts that use artificial
intelligence to steer discussions and promote specific ideas or products
on social media such as Twitter and Facebook. To typical social media
users browsing their feeds, social bots may go unnoticed as they are
designed to resemble the appearance of human users (e.g., showing a
profile photo and listing a name or location) and behave online in a
manner similar to humans (e.g., 'retweeting' or quoting others' posts
and 'liking' or endorsing others tweets).}'' 

However, as has been pointed out by Rauchfleisch and Kaiser~\cite{rauchfleisch2020false}, there is some confusion as to what exactly ``social bot'' researchers or
tools like \Botometer\ are trying to find. Authors in this field often use the terms `social bot', `bot', `social media bot', or `automated account' more or less interchangeably, even though---according to the above definition---automation is a necessary but not a sufficient condition for a \socialbot. A deeper discussion of these issues can be found in~\cite{gorwa2020unpacking}. For the purposes of this paper, however, 
the exact definition of these terms will not matter. In particular, we will 
demonstrate that the vast
majority of the accounts that are flagged as ``bots'' by \Botometer\ are real people and do not involve any automation at all. In the rare occasions
where we found partly automated accounts, e.g. automated retweets or accounts that automatically cross-posted content from other social media
platforms on Twitter, we will point this out explicitly.

This paper is structured as follows: In Section 2, we discuss
theoretical and methodological limitations of automatic bot-detection tools 
like \Botometer.  We point out 
a fundamental theoretical flaw of the commonly used approach of estimating the level of 
social bot activity.
In Section 3, we evaluate the performance of \Botometer\ empirically
using various samples of Twitter accounts and discuss related research
that used a similar approach. When using \Botometer\ on accounts of
known humans, the false-positive rate turns out to be significant.
In Section 4, we describe our experiments 
to evaluate the validity of \Botometer\ scores in real-world scenarios.
To our knowledge, a systematic evaluation
of this type has never been reported before. Our results are devastating for the whole body of \Botometer-based research: Nearly all accounts that are labeled as ``bots'' based on \Botometer\ scores are false positives. Many of these accounts are operated by people
 with impressive academic and professional credentials. Not a single one
 of the hundreds of accounts we inspected---each of which had been
 flagged by \Botometer---was a ``social bot'' according to the above
 definition.  In Section 5, we present our conclusions.

\section{Theoretical and methodological limitations of Botometer-based \socialbot\ detection}

\Botometer\ is an automated tool designed to discriminate
\socialbots\ from human users. It is built on a supervised machine
learning approach. To discriminate between the two classes, a random forest classifier is 
trained on two samples of user accounts, one labeled ``human'' and one
labeled ``bot''. The classification is based on more that 1,000
features which, according to~\cite{davis2016botornot}, include statistical features of retweet networks, 
meta-data, such as account creation time, the median number of followers of an accounts social contacts, the tweet rate, and features based on part-of-speech tagging and sentiment analysis.

The training of \Botometer\ is based on a publicly available
dataset\footnote{\url{https://botometer.osome.iu.edu/bot-repository/datasets.html}}
where (in 2019) 57,155 accounts were labeled ``bot'' and
30,853 were labeled ``human''~\cite{yang2019arming} The accounts come
from a variety of different sources and many of the labels seem at least
questionable. The largest subset of ``bots'' comes from a sample
of spammy or promotional accounts from the early days of Twitter (2009 - 2010). The
study where these accounts were collected referred to them as ``content
polluters'' and
did not claim that these accounts were automated or
bots~\cite{lee2011seven}. Many of the accounts from other sources
were apparently labeled manually by
laypersons with little understanding of the state-of-the-art in
human-machine interaction and the difficulty
of evading Twitter's detection of nefarious platform use, and based on
a na\"{\i}ve understanding of what constitutes a
``bot'' (possibly based on questionable clues like a high amount of retweets, a small or large number of
followers, missing profile picture, digits in the Twitter handle, or, as
empirically validated in~\cite{wischnewski2021disagree}, 
opposing political views). Some accounts in the ``bot repository'' were
explicitly labeled as ``bots''
because they appeared to have  participated in ``follow trains'', a technique used
by human political activists on Twitter to rapidly increase their
follower count. Clearly, the lack of reliable ground truth data is the first
glaring methodological problem of \Botometer. It seems far from obvious
that training a classifier on a rather arbitrary
selection of account samples which are based on vastly different ideas of what
constitutes a ``bot'' will result in a useful tool.

\Botometer\ is available both over an API and over a web interface. It
provides a score between 0 and 1 for individual Twitter accounts which
is calculated by calibrating the
raw score provided by the random forest classifier. Higher scores are
associated with a higher ``bot likelihood'' on accounts that are labeled
``bot'' in the bot repository. This ``bot likelihood'' is
linearly rescaled to a scale from 0 to 5 and presented on the website.
Additionally, a ``complete automation probability''
(CAP) is provided since version 3. The CAP is based on a non-linear rescaling of the
bot score according to the Bayes rule and is supposed to be interpreted as the 
posterior probability that an account is a bot. According
to~\cite{yang2019arming}, the CAP is based on the assumption that the
prior probability of observing a bot is 0.15 and provides ``generally more
conservative'' scores than the original bot score. That is, the
rescaled ``bot likelihood'' is based on the assumption that roughly 50\% of the accounts
encountered by \Botometer\ are bots, whereas the CAP is based on the
assumption that 15\% of these accounts  are bots.\footnote{
Notably, this 15\%
estimate was obtained using an earlier version of \Botometer\ as a classifier and without
manually verifying those results.~\cite{varol2017online}}

Now consider the typical methodology of most disinformation studies
which employ \Botometer\ (or similar tools). Two of many such studies
will be discussed in detail in Section~\ref{sec:realworld}:
\begin{enumerate}
\item A large sample of tweets or user accounts is collected related to a certain topic, for example, all followers of certain accounts or all tweets that contain certain hashtags or keywords, e.g. `political issue'.
\item The list of accounts is fed into \Botometer\ and the resulting ``bot scores'' are stored in a file.
\item The study authors take a look at the histogram of the ``bot
	scores''. On this basis, a suitable threshold is selected in
		some obscure or arbitrary manner. This step is often skipped and instead, a threshold of 50\% (2.5 out of 5) is employed.
\item The amount of ``bots'' and ``bot-generated'' tweets is calculated based on this threshold, resulting in headlines like ``{\it 30\% of the Twitter users who tweet about `political issue' are bots}'', or ``{\em  half of the tweets about `political issue' are generated by bots}''.
\end{enumerate}
Even under the optimistic assumption that there is a significant
correlation between the ``bot score'' and the true nature of the account, this approach is fundamentally flawed. By adjusting the threshold, almost any amount of
``bots'' between 0\% and 100\% that is desired or expected by the authors can be produced as a result (see Figure~\ref{fig:botscoredistribution}). Researchers expecting a 
large number of \socialbots\ will 
choose a low threshold, while researchers expecting a low number of \socialbots\ will choose a high threshold.\footnote{This resembles a technique of adjusting a bedridden patient's blood pressure reading by heavily tilting the bed, as described in Samuel Shem's satirical novel {\em House of God}: ``{\em You can get any blood pressure you want out of your gomer.}''}  On the other hand, if the threshold is not adjusted, the implicit assumption is that the relative number of bots in the real-world sample is the same as in the training and/or validation data used to optimize \Botometer. In either case, using this approach to estimate the prevalence of ``bots'' is a textbook example of circular reasoning. The prevalence of \socialbots\ desired or expected by the researchers directly affects the prevalence that will be ``measured''. 

The problem can also be pointed out in a probabilistic framework. An optimal classifier for ``bots'' vs. ``humans'' will follow the 
Bayes decision rule which
minimizes the number of errors. Therefore, based on a feature vector $\vec{x}$, it will make a decision for the class {\it Bot} if and only if
\begin{equation}
 p(Bot) \cdot p(\vec{x} | Bot) > (1 - p(Bot)) \cdot p(\vec{x} | Human).
\end{equation}
\Botometer, like any other classifier, incorporates an estimate of the
prior probability $p(Bot)$, explicitly or implicitly. As discussed
above, the calibrated ```bot score'' is based on the assumption that
$p(Bot)$ is roughly 50\%, whereas the CAP is based on the
assumption that $p(Bot)$ is 15\%.  Adjusting the
 ``bot score'' threshold is equivalent to modifying the prior probability $p(Bot)$.
However, $p(Bot)$ is nothing but the prevalence of \socialbots\ that researchers are trying to measure in the first place. This is a circular dependency: In order to classify accounts into ``bots'' and ``humans'' with the goal of using the relative frequency of ``bots''
as an estimate for  $p(Bot)$, we already need to know $p(Bot)$ beforehand.

Using a classifier like \Botometer\ to 
obtain a reliable estimate of $p(Bot)$ by counting unvalidated classification results would require probability density 
functions $p(\vec{x} | Bot)$ and $p(\vec{x} | Human)$ with virtually
no overlap. Only then, the impact of the unknown probability $p(Bot)$ on the classification result could reasonably be neglected. However, looking at actual ``bot score''
distributions and given the consistent failure of \Botometer\ in our experiments, this is clearly not the case.\footnote{A similar problem arises 
when human labelers are instructed to rate accounts as ``bots'' or ``humans''. The ratings will typically be based on unrealistically high expectations of the bot prevalence $p(Bot)$ (fueled by \Botometer-based publications and media coverage), a  limited
understanding of the state of the art in artificial intelligence combined with misconceptions of what features
might be ``bot-like'' (i.e. a bad estimate of $p(\vec{x} | Bot)$), as well as false and narrow expectations of what
a ``normal'' human behavior on Twitter might be (i.e. a bad estimate of
$p(\vec{x} | Human)$~). As a result, many accounts that are clearly not
automated but were rated ``bots'' by human labelers can be found in the
``bot repository'' used to train \Botometer.}

These considerations leave little hope that \Botometer\ or similar tools
might be of {\em any} value for estimating the prevalence of ``bots''.

\section{Evaluating Botometer on samples of known humans}

The problem that \Botometer\ produces enormous amounts of false
positives and that it should never be trusted without manual verification has
been pointed out for years~\cite{kreil2017social,gallwitz2019fairy,kreil2019army,kreil2020people,rauchfleisch2020false}.

\begin{figure}
\centering
\includegraphics[scale=0.143]{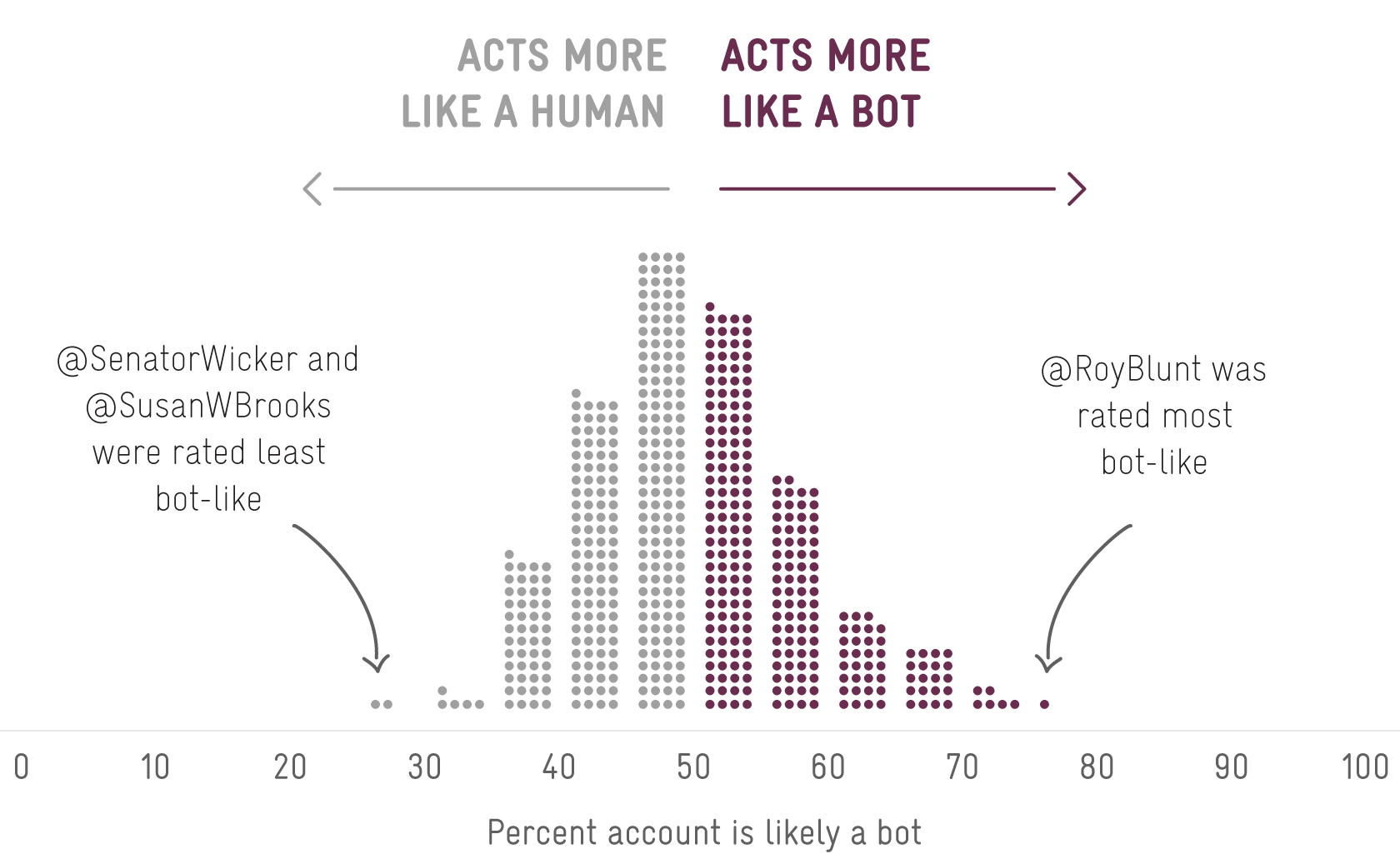}
\caption{Distribution of the \Botometer\ ``bot scores'' for the members of the U.S. Congress who were present on Twitter (April 2018)
	} 
\label{fig:congress}
\end{figure}
Different methods have been used to demonstrate the problem. A simple
and effective way is to use \Botometer\ to classify accounts that are
without doubt operated by humans. When we tested \Botometer\ in April
2018, nearly half of U.S. Congress members present on Twitter were
misclassified as bots (47\%), using the most commonly used ``bot score''
threshold of 50\% (or 2.5 on a scale from 0 to 5), see
Figure~\ref{fig:congress}. 
\footnote{Surprisingly, in May 2019, \Botometer\ performed dramatically better on the members of Congress; 
the false positive rate dropped from 47\% to 0.4\%. Possibly, these accounts had been
added to the  \Botometer\ training data as examples of human users in
	the meantime.} 

In similar experiments in May 2019
\cite{gallwitz2019fairy,kreil2019army,kreil2020people}, we
found that
\begin{itemize}
\item 10.5\% of NASA-related accounts are misclassified as bots. 
\item 12\% of Nobel Prize Laureates are misclassified as bots. 
\item 14\% of female directors are misclassified as bots. 
\item 17.7\% of Reuters journalists are misclassified as bots.
\item 21.9\% of staff members of UN Women are misclassified as bots. 
\item 35.9\% of the staff of German news agency ``dpa'' are misclassified as bots.
\end{itemize}

\begin{figure}
\centering
\includegraphics[scale=0.102]{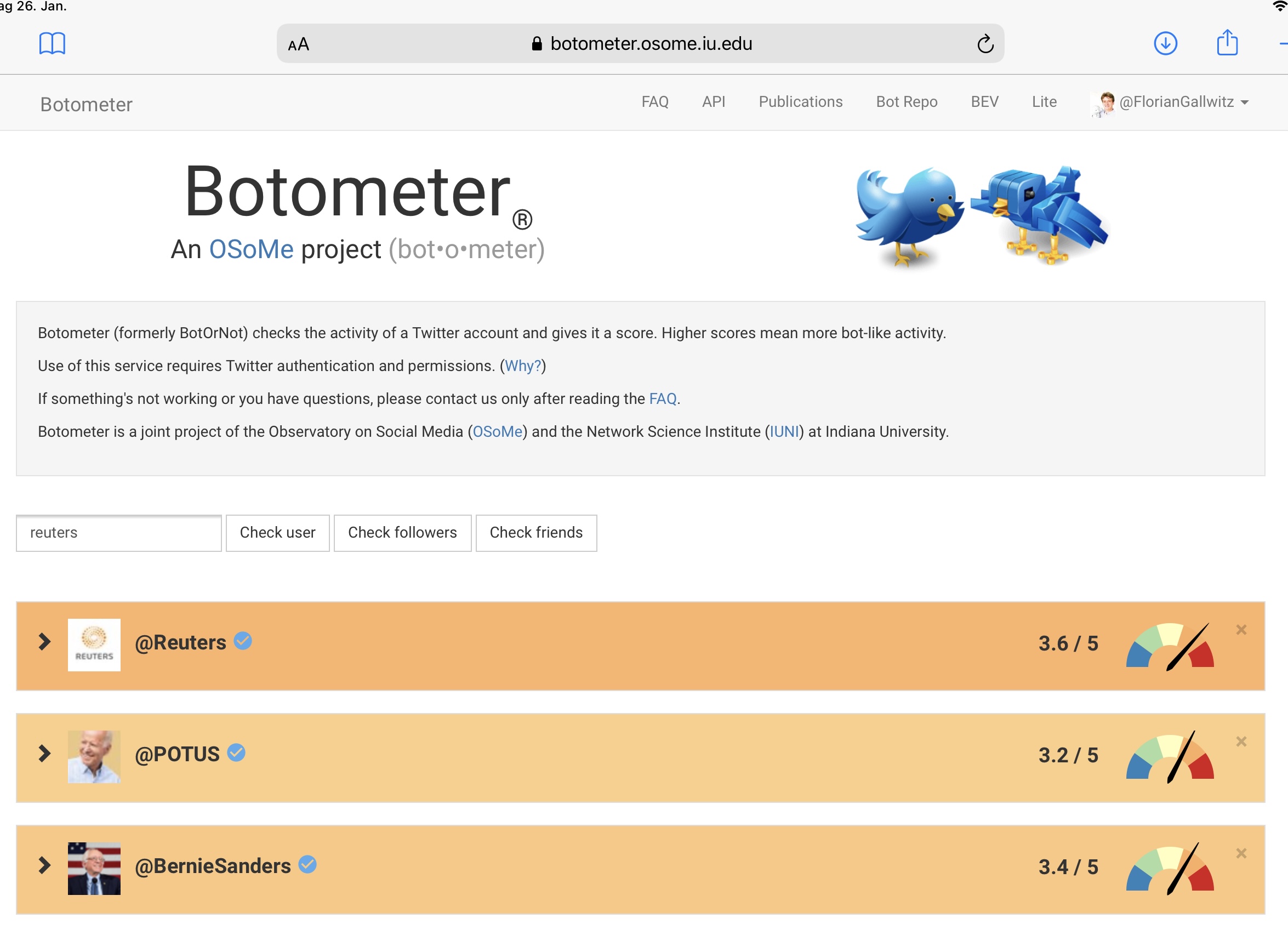}
\caption{Screenshot of \Botometer's web interface showing the ``bot scores'' of the verified Twitter accounts @Reuters, @POTUS, and @BernieSanders, each being misclassified as a bot (January 2021)}
\label{fig:botometer}
\end{figure}

Clearly, \Botometer's glaring false positive problem has still not been solved with the current version. In 
January 2021, even the Twitter account @POTUS of the newly elected president of the United States, Joe Biden, was classified 
as a ``bot'', with a ``bot score'' of 3.2, see
Figure~\ref{fig:botometer}. In May 2022, both presidential accounts @POTUS and @JoeBiden received ``bot scores'' of 3.8.

The lack of reliability goes both ways. When we tested \Botometer\ with real, automated Twitter bots in May 2019, we found that
\begin{itemize}
\item 36\% of known bots by New Scientist are misclassified as humans.
\item 60.7\% of the bots collected by Botwiki are misclassified as humans. 
\end{itemize}
A similar, but more systematic approach to evaluate the ability of  \Botometer\ to discriminate between automated accounts and humans was chosen by~\cite{rauchfleisch2020false}. The authors performed experiments on five datasets of verified bots and verified humans. Although the datasets had been partly used to train \Botometer, the authors find that ``{\em the \Botometer\ scores are imprecise when it comes to estimating bots. [...] This has immediate consequences for academic research as most studies using the tool will unknowingly count a high number of human users as bots and vice versa.}''

Although they clearly demonstrate severe limitations of \Botometer, evaluations on manually selected lists of accounts do not allow for an estimate of \Botometer's reliability in a real-world setting. In a real-world scenario, an enormous variety of human user behaviors may occur which might not be included in manually constructed test samples. Also, it seems unlikely that actual malicious ``social bots'' would behave in the same manner as the bots employed in these experiments.
Therefore, the only way to get a realistic impression of  \Botometer's performance in real-world scenarios is to take a closer 
look at the accounts that are classified as bots in a specific setting. 
Unfortunately,  it turns out that authors of studies of this type are extremely reluctant to share their lists of ``bots''. In our impression, most of the study authors we contacted were fully aware that the lion's share or even all of the accounts they counted as ``bots'' were, in fact, humans.

This reluctance to share data motivated us to replicate one study of
this type about alleged \socialbots\ among the followers of German
political parties~\cite{keller2019social}. In only one case, we were
able to obtain the relevant raw data from the authors of a peer-reviewed
\Botometer-based study. We are grateful that Dunn et. al.~\cite{dunn2020limited} shared their list of ``bots'' with us to allow us to take a closer look. 

\section{Evaluating the performance of Botometer in real-world
scenarios}\label{sec:realworld}

In this section, we will first evaluate the performance of \Botometer\
on the basis of two peer-reviewed studies where this tool was employed
to estimate the prevalence of \socialbots. Lastly, we will summarize the
findings of two recent studies where  \Botometer\ results have been
checked manually. 

\subsection{Are \socialbots\ following the Twitter accounts of German political parties?}

Keller and Klinger~\cite{keller2019social} analyzed Twitter data that was collected before and during the 2017 federal election campaign in Germany. Based on an analysis with \Botometer, they claim that among the followers of seven political parties, ``{\em the share of social bots increased from 7.1\% to 9.9\% during the election campaign}''. The total numbers of Twitter followers they analyzed during the election campaign was 838,026, which, assuming the claimed \socialbot\ prevalence of 9.9\%, would correspond to roughly 83k \socialbots. 
However, the paper does not provide any examples of \socialbot\ accounts and the raw data was not shared. When we contacted the authors, they were unable to provide us with a single credible example of a ``social bot''.

In order to verify the validity of these results, we tried to replicate Keller's and Klinger's approach in May 2019. Although we expected some changes in the follower lists of the political parties over the 20 month period between September 2017 to May 2019, we see no reason for (or evidence of) a fundamental change.

When we downloaded the followers of the seven political parties using the Twitter API, we found a total of 521,991 different accounts that had at least tweeted once (\Botometer\ is not able to provide scores for accounts without tweets). To classify these accounts into ``bots'' and ``humans'', we used the \Botometer\ API to determine the ``bot score'' for each of the accounts and used the same, unusually high ``bot score'' threshold that Keller and Klinger had chosen for their study: 76\%, or 3.8 on a scale from 0 to 5.

\begin{figure}
\centering
\includegraphics[scale=0.325]{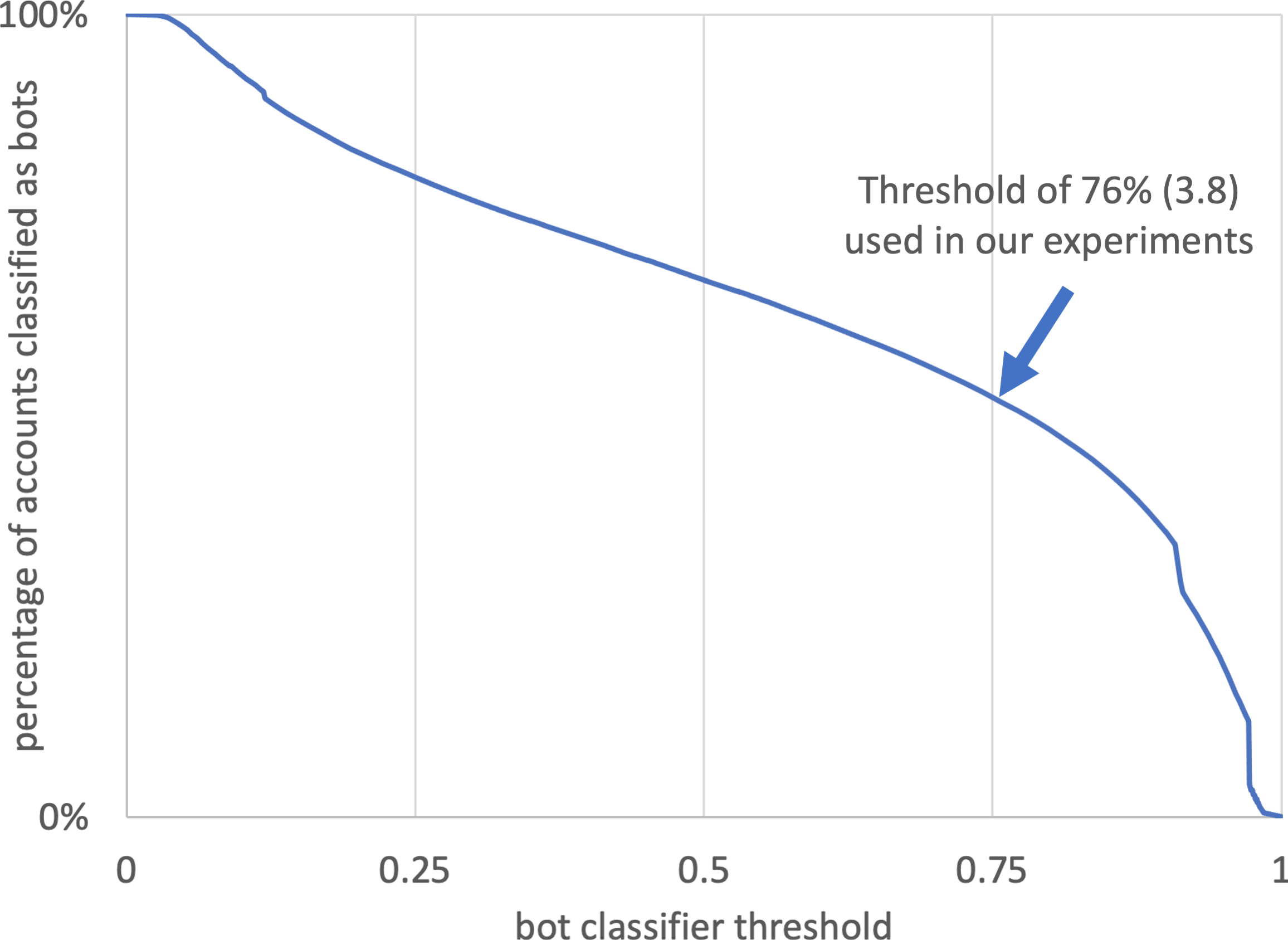}
\caption{Cumulative distribution of the \Botometer\ ``bot scores'' for the Twitter accounts following the accounts of
the largest German political parties
in May, 2019. The threshold of 76\% (3.8 out of 5) which was defined by \cite{keller2019social} and reused in our replication attempt is highlighted. }
\label{fig:botscoredistribution}
\end{figure}

Surprisingly, the amount of \socialbots\ appeared to have increased dramatically since the elections. We found a total of 270,572 accounts that exceeded the ``bot score'' threshold of 3.8. This corresponds to a \socialbot\ prevalence of 51.8\%. Mysteriously, the amount of \socialbots\ among the followers of German political parties appeared to have increased fivefold in the 20 months since the election, while the total number of followers had decreased.  The commonly used threshold of 50\% or 2.5 would have resulted in a ``social bot'' prevalence of 67\%, see Figure~\ref{fig:botscoredistribution}.

In order to understand what was really going on, we chose to take a
closer look at the list of \socialbots. Obviously, it is not feasible to
manually analyze 270k Twitter accounts. In order to assess the true
nature of these accounts, we decided to select a sample of slightly more
than 100 of the alleged ``social bot'' accounts. While random sampling
might have been feasible, we preferred a deterministic strategy to make
our results reproducible. At the same time, the approach should
guarantee a representative sample.  For this purpose, we sorted the 270k
``bot'' accounts in descending order of their ``bot score''. Within that
list, accounts with the same ``bot score'' were sorted in descending
alphabetical order according to their Twitter handle. From the resulting
list, we selected all accounts in lines where $linenumber$ mod $2,500 = 1$, i.e. the 1st, 2501st, 5001st, ... account. This selection procedure resulted in a list of 109 alleged \socialbots\ covering the range of ``bot scores'' from 3.857 to 4.931.

Each of the 109 accounts was closely analyzed manually, using tools like https://accountanalysis.app to check for 
unusual timing patterns and to retrieve the Twitter clients that were used for recent tweets and retweets. Most importantly, we closely 
inspected the tweets and retweets as well as the interactions with other users to search for traces of potential automation.

The complete list of accounts with the ``bot score'' assigned by \Botometer, the Twitter handle, the follower number, and a short comment about distinctive 
characteristics of the account can be found in the Supplemental
Materials, Section
A.\footnote{\url{https://www.in.th-nuernberg.de/Professors/Gallwitz/gk-md22-suppl.pdf}} Here, we want to give some examples of such accounts. We will also speculate about potential reasons why these accounts might have been misclassified as ``bots''.
\begin{itemize}
 \item More than 20\% of the accounts that were misclassified as ``bots'' only tweeted a single tweet after the account was created. We found this to be a reproducible behavior of \Botometer: After creating a new account and tweeting a single tweet, any account is rated 
 a ``bot'' with a close-to-maximum ``bot score''. A somewhat tragic
		example of this type is a  Twitter user in our sample, a
		follower of the conservative party account @CDU, who
		asked @Microsoft for help regarding a problem with her
		Microsoft account in July 2017 and (as of May 2022) has never received a reply. 
 \item Inactivity in the last couple of months or years seems to be
	 another main reason for \Botometer\ to rate accounts as
		``bots''. This is a quite surprising behavior, given the
		common narrative that \socialbots\ are supposed to be
		highly active accounts, as suggested by the Oxford
		50-tweets-per-day criterion (see, for example,
		\cite{howard2016elec}). While \Botometer\ still displays
		this behavior in May 2022, a warning has been added to
		the web interface in the meantime: {\em @accountname is not active, score might be inaccurate}.  
 \item Although quite active, the Twitter account @CDU\_VS, the local
	 branch of the conservative party in the town of
		Villingen-Schwenningen, is rated a ``bot''. A possible
		reason seems to be that the operator of the account is
		quite frequently cross-posting content from their
		Facebook and Instagram pages directly through the
		``Tweet'' button on these platforms. However, this is a perfectly normal, frequent and acceptable way of sending tweets and has nothing to do with the concept of a \socialbot. User @hoockstar, for example, also might have turned himself into a ``bot'' by posting a couple of Instagram links in this manner.
 \item A similar issue might have turned six other Twitter users in our
	 sample into ``bots''. Each of them sent at least one tweet through a game app, such as ``8 ball pool'' or ``ClumsyNinja''. Tweeting your progress in games like these is often honored with a reward in some kind of in-game currency, while the tweet serves as an ad for the creators of the game.
 \end{itemize}
To summarize the result, not a single one of the accounts in the list
was a ``bot'' in any meaningful sense of the word, certainly not a
\socialbot\ according to the definition given in the Introduction. In
other words, every single ``bot'' in our sample was a false alarm.
Assuming that \Botometer\ does not perform {\em worse} than a random
number generator on the followers of German political parties (as we did not check the accounts rated as human), we can also conclude that the best guess for the number of \socialbots\ following the political parties in Germany is not 83,000, as claimed by Keller and Klinger, but zero.

\subsection{Are \socialbots\ attempting to spread vaccine-critical information?}\label{sec:vaccine}

The second study we used as a basis for evaluating \Botometer\ in a real-life scenario investigates the influence of ``bots'' in the spread of vaccine-critical information \cite{dunn2020limited}. The authors examined a study population of 53,188 U.S. Twitter users who were selected independently of whether they were exposed to or shared vaccine-related tweets or not.  Additionally, about 21m vaccine-related tweets coming from approx. 5m accounts had been identified based on keyword filtering. The study examined whether and how the users of the study population interacted with vaccine-related tweets, and whether the vaccine-related tweets users were exposed to came from ``bots''. 

In order to determine if a Twitter account was a bot, \Botometer\ was employed, using the usual threshold of 50\% (2.5 out of 5). A total of 5,124,906 accounts was scored, resulting in 197,971 ``bots'', which corresponds to a ``bot'' prevalence of 3,8\%.

Although the study concluded that the exposure to bot-generated content was limited---which seemed highly plausible to us---we were skeptical about the true nature of the accounts that had been counted as ``bots''. 

The authors kindly provided us with the list of Twitter UserIDs that were counted as ``bots'', as well as the corresponding \Botometer\ scores. The scenario differs from the German party follower study in important aspects which should help to avoid two main causes
of false positives: 
\begin{itemize}
\item The sample of 5m accounts was selected based on active tweets that contain vaccine-relevant keywords. Therefore, inactive accounts (which, as we have seen, are commonly misclassified as ``bots'' by \Botometer) could not become part of the sample.
\item Given that every account in the sample had to post at least one tweet with a vaccine-related keyword, it was highly unlikely (though, as it turned out, not impossible) that accounts which had only produced a single tweet in their lifetime could become part of the sample.
(Accounts like these, as we have seen, are commonly---if not always---misclassified as ``bots'' by \Botometer.) 
\end{itemize}
Therefore, the dramatically lower prevalence of ``bots'' in this sample (3.8\% vs. 67\% on the basis of the 50\% threshold) is not surprising. 

Again, we wanted to select a suitable, representative subset of the 197,971 ``bot'' accounts in a deterministic manner which would undergo manual analysis. The list
provided to us by~\cite{dunn2020limited} included unique numerical user IDs, not the alphanumeric Twitter handles we had used in the party follower scenario. 
For deleted or suspended accounts, we could not
reconstruct the alphanumeric Twitter handles. Also, Twitter handles may
change over time. Therefore, in order to keep our approach reproducible,
we used a selection strategy which was directly based on the numerical
user IDs: We sorted the list of accounts numerically in increasing order
according to their unique Twitter UserID and selected all accounts in
lines where $linenumber$ mod $1,500 = 1$, i.e. the 1st, 1,501st, 3,001st, ... account. This resulted in a list of
132 alleged \socialbots\ covering the range of ``bot scores'' from 2.557 to 4.871.

Of these 132 accounts, 11 had been deleted since the data had been collected (2017-2019).
On the basis of the numerical Twitter UserID alone, without access to the Twitter handle, an investigation through the Internet Archive was not successful. This left us with 121 ``bot'' accounts that we could analyze.
 
 We used tools like https://accountanalysis.app to check for 
unusual timing patterns and to retrieve the Twitter clients that had been used for recent tweets and retweets. Most importantly, 
we closely inspected the tweets and retweets as well as the interactions with other users to search for traces of potential automation. 
We put a significant amount of effort into establishing the true identity of the persons behind the accounts wherever possible, using Google Search, 
Facebook, LinkedIn, photo comparisons, homepages of physicians and scientific institutions, as well as various other information sources 
available on the internet.


The complete list of accounts with the ``bot score'' assigned by \Botometer, the Twitter handle, the follower number, and a short comment about distinctive 
characteristics of the account can be found in the Supplemental
Materials, Section
B.\footnote{\url{https://www.in.th-nuernberg.de/Professors/Gallwitz/gk-md22-suppl.pdf}}

A surprisingly high number of the 121 alleged ``bots'' in our sample are
in reality individuals with academic or professional credentials, many
of them directly related to the topic of vaccines.
Each of them used their real name in their Twitter bio. The
list included (in the order of User ID) a medical intern
and researcher from Saudi Arabia, an
International Development and Public Health professional at the
University of London, a pediatric nurse practitioner in New York, a
senior lecturer at the School of Business at the \"{O}rebro University,
a Technical Manager at a health products company in Nigeria, an IT
professional in Tohana, India, a former Pediatrician who is now working
at the Embassy Medical Centres of HOPE Worldwide in Cambodia, a human
resource professional in Michigan, a specialist in trauma surgery at
Oscar G Johnson Veterans Admin. Hospital, the former Director of WHO's
Prevention of Noncommunicable Diseases who holds multiple academic
degrees and titles, a pediatrician in Loma Linda, California, a medical
student at the University of Rwanda, a postdoctoral researcher at the
Institut de Recherche en Infectiologie de Montpellier, a junior
researcher working in the area of infectious diseases at the University
Medical Center Rotterdam, an Associate Professor at the University of
Texas Southwestern Medical Center and a Senior Economist at the RAND
Corporation, a student in the Master's program Health Studies at the
Athabasca University in Canada, and a Public Health Officer from Kampala
University.
 
\begin{figure}[tb]
\centering
\includegraphics[scale=0.1]{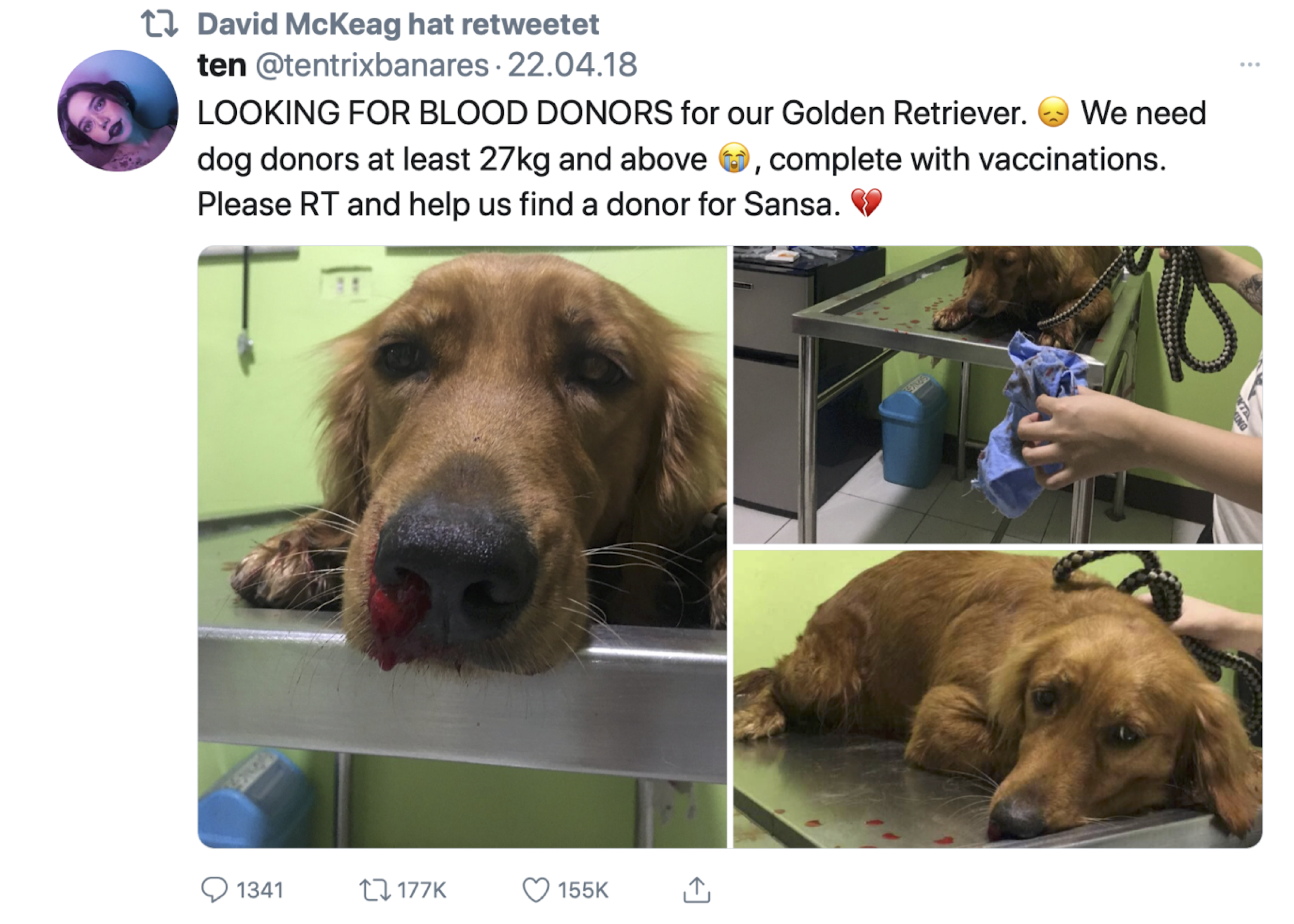}
\caption{Viral tweet about a dog in need of a canine blood donor. As it
	 contains the keyword `vaccinations', thousands of users who had
	 retweeted this tweet and were
	 misclassified as ``bots'' by \Botometer\ were counted as
	 ``vaccine bots'' in~\cite{dunn2020limited}.}
\label{fig:retriever}
\end{figure}

 A number of the 121 alleged ``bots'' are, in reality, the official Twitter accounts of health-related organizations:
 \begin{itemize}
 \item @THEWAML, The World Association for Medical Law
 \item @CdnAcadHistPhm, The Canadian Academy of the History of Pharmacy
 \item @Infprevention, The Infection Prevention and Control Conference IPCC
 \item @jubileetanzania, The Jubilee Insurance Company of Tanzania Ltd.
 \item @JFoundation\_, The Jamachiz foundation in Lagos which supports the improvement of basic human welfare.
 \item @UNMCkidney, The University of Nebraska Medical Center Division of Nephrology 
 \end{itemize}
 Animals are commonly vaccinated as well. This contributed to the fact that some of the alleged ``vaccine bots'' are, in reality, Twitter accounts related to agricultural topics and pets:
 \begin{itemize}
 \item @agriview, ``Wisconsin's Leading Agriculture Newspaper''
 \item @top5stories, Links to stories and videos about pitbulls, e.g. americanbullydaily.com
 \item @Thinkagro, Thinkagro Co Ltd., a China-based provider for agricultural products and solutions
 \item At least two of the accounts ended up as ``vaccine bots''
	 because, like 177k other users, they had retweeted a tweet
		 about a seriously ill Golden Retriever in need of a
		 canine blood donor. The tweet included the keyword
		 `vaccinations', see Figure~\ref{fig:retriever}.
 \item @anonymousinapp1 tweeted only a single tweet with a joke that
	 contained the word `vaccines` and got one like for it.
 \end{itemize}

Some of the accounts in the list use a certain degree of automation, none of which has anything to do with attempts to spread vaccine-critical information:
\begin{itemize}
\item A hospital doctor used the RoundTeam Twitter content management
	platform to tweet about hospital-related topics. The commercial
		service claims that it ``{\it enables you to grow your
		presence on Twitter}''. (The doctor who used this service
		for a year acquired a total of 11 followers---highlighting how difficult it is to acquire followers even
		for a real and reputed person if their tweets remain
		generic and impersonal.)
\item @Onderzoekers is the automated Twitter feed of a Dutch website
	which offers job openings in science and research.
\item @top5stories uses the web-based automation tool IFTTT to automatically posts links to newly published articles on a some pitbull related sites.
\item @vectorborg is a hashtag retweet bot with 1k followers that
	retweets tweets which contain the hashtag \#GreenEnergy. Retweet
		bots of this type were fairly common in the early days
		of Twitter.  For example, a list of ''{\em 77 Useful
		Twitter Retweet Bots}'', including dozens of bots
		dedicated to retweeting tweets that mention a German
		city from \#Aachen to \#Wuerzburg, was presented
		by~\cite{dreissel2010retweet}. However, retweet bots
		give spammers easy access to the timeline of the
		followers of these bots, simply by including hashtags or
		keywords in unrelated tweets. As of May 2022, most retweet bots---including the ones in the list---have been deactivated or suspended by Twitter.
\item @LoydGailpg is an account with 6 followers that used IFTTT to retweet links to newly published airline and travel related articles.
\end{itemize}
To summarize, in a representative sample of 121 of the almost 198k accounts that were counted as ``bots'' in~\cite{dunn2020limited}, we found 116 human-operated accounts with no signs of automation. We can therefore estimate that approx. 190k human accounts
were falsely counted as ``bots'' in this study. Many of these Twitter users have impressive academic and professional credentials. 
Consider that approx. 1,500 scientists and experts with similar credentials have been misclassified as \socialbots\ for each of the scientists and 
experts in 
our sample. Only 5 of the accounts in our sample might be considered (unmalicious) bots. Not a single one of these accounts had anything to do with automated attempts at spreading vaccine-critical information or disinformation. 

Again, we could not find a single account that fits the usual definition of a \socialbot\ as cited in the Introduction. 

\subsection{Further evidence for the failure of Botometer in real-world scenarios}

Our devastating findings about \Botometer's lack of reliability in
real-world scenarios are consistent with findings in two recently published
studies.

In~\cite{he2021people}, the Twitter debate around
mask-wearing during the COVID-19 pandemic was analyzed. Like so many disinformation researchers before
them, the authors used \Botometer\ to detect the ``social bots'' in
their sample of Twitter accounts. However, they followed the
recommendations in~\cite{rauchfleisch2020false} and manually labeled a random sample of 500 distinct users to check the reliability of \Botometer's results.
Similar to our approach, they took a look at user profiles, tweeting histories, and interactions with other users in order to determine 
whether they were dealing with a \socialbot\ or not. 

They did not find a single bot in their sample of 500 accounts. \Botometer, however, labeled 29 (or 5.8\%) of these accounts as ``bots''.  

The authors decided to ignore the \Botometer\ results: ``{\em [...] based on manual review, many of these users were simply hyper-active tweeters. Their online activities did exhibit the normal behavior of human users, e.g. the content they posted did not appear to be automatically authored and they participated in active interactions with other Twitter users.}''
 
In~\cite{poddar2021winds}, the vaccine-related stance of Twitter users
during the COVID-19 pandemic was investigated. As the authors wanted to focus on
human users, they tried to exclude bots using Botometer. In their
sample of 675 accounts, 68 received a score (presumably the CAP) of 0.5 or higher.
However, the authors also followed the recommendations
by~\cite{rauchfleisch2020false} and manually checked these accounts.
They found only one of these accounts to be automated, an account
that shared articles from a personal blog on Twitter. The other 67
accounts were human users, i.e. false alarms.

\section{Conclusion}

The field of \socialbot\ research is fundamentally flawed.  While \socialbot\ researchers have received an enormous amount
of public attention, the vast majority of their findings fully rely on the accuracy
of \Botometer.  Many researchers simply refer to all accounts in their
studies as ``bots'' or
even as ```social bots'' as long as they exceed some arbitrarily chosen \Botometer\
threshold. However, this approach is highly questionable from a 
theoretical point of view. And, as we have
demonstrated empirically, Botometer fails miserably and consistently when evaluated under real-world conditions. 
Studies claiming to investigate the prevalence, properties, or influence
of \socialbots\ based on \Botometer\ have, in reality, just investigated
false positives and artifacts of this approach. 

While automated Twitter accounts exist, we have yet to see a single
credible example of a malicious \socialbot, i.e. an account that pretends to be a human
user and is operated by some sinister actor to manipulate public
opinion. 

In our impression, a prevailing 
culture of intransparency is a major factor that has enabled an entire
field of research to rely on deeply flawed methods. Publishing or sharing raw data, as it is common practice in many other fields of science, would have helped to identify and highlight the fundamental methodological problems much earlier.  

We conclude that all past and future claims about the prevalence or influence of \socialbots\ that are not accompanied by lists of account 
IDs are highly questionable and should be ignored. In those cases where actual \socialbot\  accounts are named,
we recommend an in-depth analysis of these accounts to verify whether these are not simply human beings---made of flesh and blood---who have been misclassified as \socialbots\ like hundreds of thousands of Twitter users before them.

\subsubsection*{Acknowledgements}
We are grateful to Adam Dunn for sharing with us the relevant raw data we used in Section~\ref{sec:vaccine}. We sincerely appreciate the valuable comments and suggestions by Adrian Rauchfleisch, Darius Kazemi, and J{\"u}rgen Hermes, which helped us to improve the 
quality of the manuscript. 



\bibliographystyle{splncs04}
\bibliography{paper}

\end{document}